\documentclass[a4paper, 11pt]{article}

\usepackage[affil-it]{authblk}

\makeatletter
\def\@maketitle{%
  \newpage
  \null
  \vskip 2em
  \begin{center}%
  \let \footnote \thanks
    {\Large\bfseries \@title \par}%
    \vskip 1.5em
    {\normalsize
      \lineskip .5em %
      \begin{tabular}[t]{c}%
         \@author
      \end{tabular} \par}%
    \vskip 1em 
    {\normalsize \@date}%
  \end{center}%
  \par
  \vskip 1.5em} 
\makeatother

\usepackage{AdamStyle}
\usepackage{caption}

\usepackage[english]{babel}
\usepackage[utf8]{inputenc}
\usepackage{amsmath}
\usepackage{amssymb}
\usepackage{graphicx}
\usepackage[colorinlistoftodos]{todonotes}
\usepackage{soul} 

\title{\bf Bayesian approach to uncertainty quantification\\ for cerebral autoregulation index}

\author{Kevin P. O'Keeffe\\ \vspace{-0.35 cm}
{\small Senseable City Lab, Massachusetts Institute of Technology, USA}\\

\vspace{-0.35 cm}

\vspace{1cm}

\vspace{-0.35 cm}

{\rm Adam Mahdi}\\
{\small Department of Engineering Science, University of Oxford, UK}
}

\date{}

\newcommand{\ari}{{\rm ARI }}

\begin{document}

\maketitle

\vspace{-1cm}
\begin{abstract}
Cerebral autoregulation refers to the brain's ability to maintain cerebral blood flow at an approximately constant level, despite changes in arterial blood pressure. The performance of this mechanism is often assessed using a ten-scale index called the ARI (autoregulation index). Here, $0$ denotes the absence of, while $9$ denotes the strongest, autoregulation. Current methods to calculate the ARI do not typically provide error estimates. Here, we show how this can be done using a bayesian approach. We use Markov-chain Monte Carlo methods to produce a probability distribution for the ARI, which gives a natural way to estimate error.
\end{abstract}

\tableofcontents


\section{Introduction }
The brain is a extraordinary system of supply and demand. It requires a rich supply of oxygen, accounting for approximately one-fifth of the total bodily intake. This demand for oxygen is supplied by virtually constant cerebral blood flow. Interruptions to this blood flow can have dire consequences, leading to haemorrhage, embolisms, and aneurysms. In order to function properly, the brain must therefore have the ability to maintain approximately constant cerebral blood flow rate (CBFV), in spite of changes in arterial blood pressure (ABP). This ability is known as cerebral autoregulation (CA) \cite{Aaslid1989}. 

The first experimental demonstration of CA was performed by Lassen \cite{Lassen59}, who derived the triphasic autoregulation curve from CBFV measurements in different human studies \cite{Dagal2009}. Later experiments found similar results both in animals \cite{MacKenzie76, MacKenzie79, Harper84, Dirnagl90} and humans \cite{Czosnyka2001}.  The four physiological mechanisms are generally identified: myogenic, metabolic, shear-dependent, and neurovascular regulation. However a complete understanding of these processes and their interactions is still lacking \cite{mader2015modeling}, making the theoretical study of CA an active area of research.

Over the years many mathematical methods have been developed to quantify CA. Common approaches are transfer function analysis, time-series models such as FIR (finite impulse response) and ARX (autoregressive exogenous), and various physiologically-based modeling techniques. Classic ODE models have also been used. One of the simplest and popular model-based approaches was proposed by Aaslid and Tiecks \cite{Tiecks95}. It consists of just three ODEs with three parameters. At a `black-box' level, the model takes time series of a patient's ABP and CBFV as input, and returns a number between 0 and 9. This number is the called the autoregulation index (ARI). A high score means the brain is auto-regulating properly, while a low one indicates the opposite \cite{Panerai08, Panerai08b}.

In spite of its simplicity, the ARI model has several limitations. The most serious of these is the inability to provide an estimate for the error in the ARI. This limits the ARI's diagnostic utility, since a clinician has no sense of the uncertainty in a given measurement. We show here how a bayesian approach can be used to overcome this limitation. After recasting the model is a more convenient form, we use the popular Markov Chain Monto Carlo (MCMC) method to give the ARI a statistical interpretation. Specifically, we produce a probability distribution for a \textit{single} ARI, whose width provides the desired estimate of the error in the ARI.

\section{Methods}
\subsection{Data}\label{data}
\noindent {\bf Data collection.}
The ABP and CBFV data, used  in this study, have been been previously used (Hebrew Rehabilitation Center for Aged, Boston, MA)~\cite{Lip2000}. ABP was measured noninvasively using a photoplethysmographic Finapres monitor (Ohmeda Monitoring Systems, Englewood, CO). In order to eliminate hydrostatic pressure effects, the subject's nondominant hand was supported by a sling at the level of the right atrium. The individuals were asked to breath at the rate of 15 breaths per minute to minimise the effects of respiration. Doppler ultrasonography was used to measure the changes in CBFV within the MCA. The 2 MHz probe of a portable Doppler system (MultiDop X4, DWL-Transcranial Doppler Systems Inc., Sterling, VA) was strapped over the temporal bone and locked in position with a Mueller-Moll probe fixation device to image the MCA. Flow velocity was recorded at a depth of approximately 50-65 mm and digitised and stored for analysis. For a complete description of the data collection procedure see~\cite{Lip2000}.

\medskip

\noindent {\bf Data preprocessing.}
The pulsatile ABP and CBFV were filtered using zero-phase 8th-order Butterworth  low pass  filter  with the cutoff frequency of $20\,\mathrm{Hz}$ (see \cite{Panerai2000}). Subsequently, the beginning and end of each cardiac cycle were marked by the onset of the systole using blood pressure signal. The onsets were detected using a windowed and weighted slope sum function and adaptive thresholding \cite{Zong2003}. The mean of ABP and CBFV were calculated for each detected cardiac cycle. A first order polynomial was used to interpolate the resulting beat-to-beat time series, which was followed by downsampling at $10\,\mathrm{Hz}$ to produce signals with uniform base. Examples of preprocessed data are shown in Figure~\ref{data}.

\begin{figure}[h!]
\centering
    \includegraphics[width= 0.8 \linewidth]{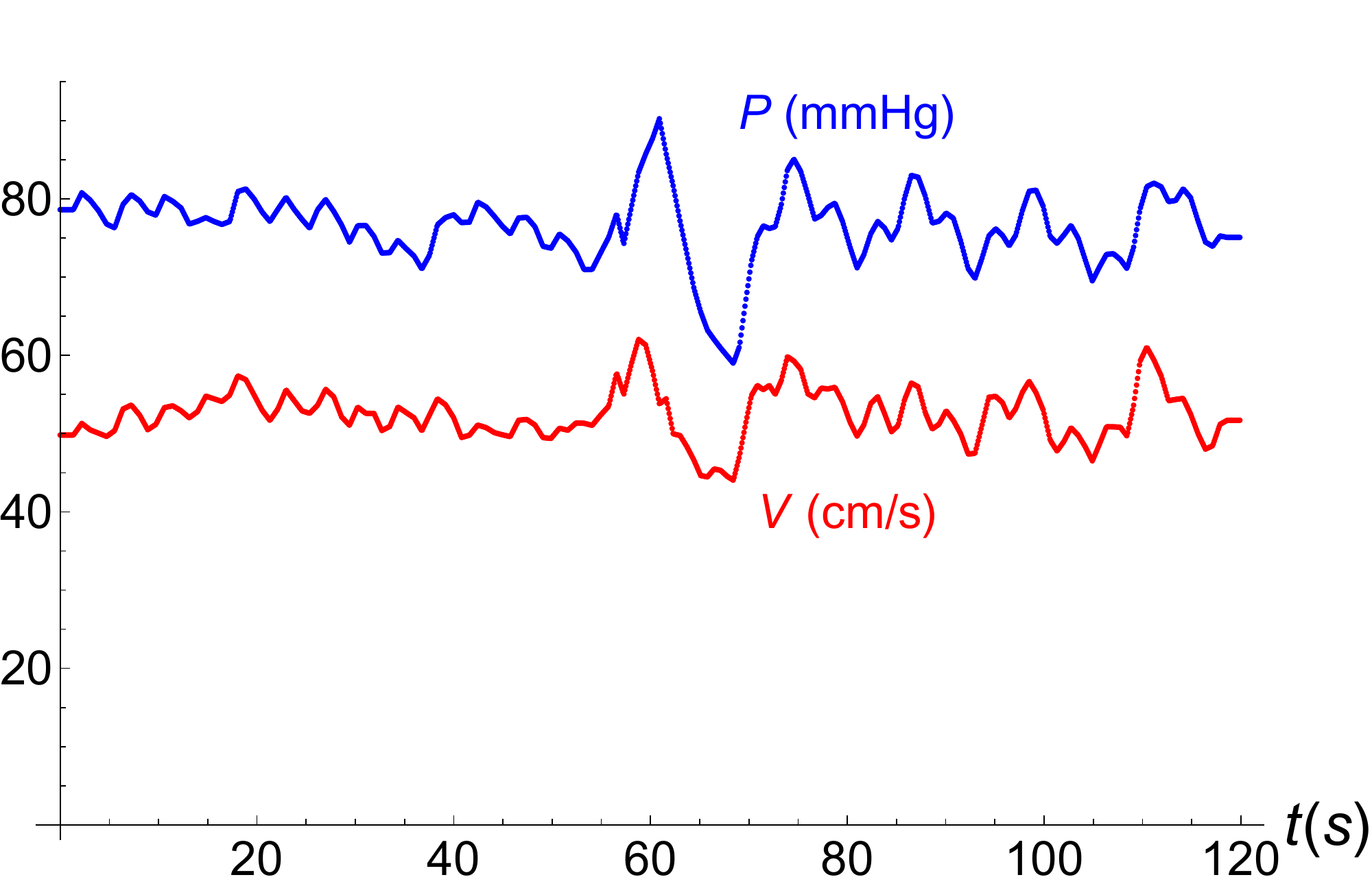}
  \caption{Time series of a patient's arterial blood pressure $P$ and cerebral blood flow velocity $V$. The data have been collected and processed as described in Section 2.1.}
  \label{data}
\end{figure}

\medskip

\subsection{Autoregulation index}

Here we describe the original Aaslid-Tiecks method \cite{Aaslid1989} to measure the ARI. We denote by $P[t]$ and $V[t]$ the two time series of ABP and CBFV, respectively, of length $N$ and indexed by $t$. Let $P_m$ and $V_m$ denote the mean value of $P[t]$ and $V[t]$ for the entire interval of interest. Initially, the time-varying ABP signal $P[t]$ is normalised as follows
\begin{equation}
dP[t]=\frac{P[t]-P_m}{P_m-P_{cr}}, \label{eq1}
\end{equation}
where $P_{cr}=12\,\mathrm{mmHg}$ is the critical closing pressure \cite{Tiecks95}. The following system of difference equations is used to compute the intermediate quantities $x_1$, $x_2$ as
\begin{equation} 
\begin{aligned}
&x_1[t]=x_1[t-1]+\frac{dP[t-1]-x_2[t-1]}{fT}\\
&x_2[t]=x_2[t-1]+\frac{x_1[t-1]-2Dx_2[t-1]}{fT} \label{eq2}
\end{aligned}
\end{equation}
where $f$, $D$ and $T$ are the sampling frequency, damping factor, and time constant parameters, respectively. The modelled CBFV, denoted by $\hat V[t]$, is computed as
\begin{equation} 
\hat V[t]=V_m (1+dP[t]-Kx_2[t]), \label{eq3}
\end{equation}
where $K$ is a parameter representing autoregulatory gain. Note that in order to start the process at steady-state the initial conditions can be selected as
\begin{equation}\label{ARI:ic}
x_1(0) = 2D\,dP[0],\qquad x_2(0) = dP[0].
\end{equation}

\smallskip

\noindent{\bf CA assessment.} Following \cite{Tiecks95}, in Table\,\ref{table:tdk} we list a combinations of ten different values of $(T,D,K)$ are used to generate ten models corresponding to various grades of autoregulation, ranging from 0 (absence of autoregulation) to 9 (strongest autoregulation). 

\begin{center}
\begin{tabular}{|c|c|c|c|c|c|c|c|c|c|c|}
\hline
 \textbf{ARI} & \bf{0} & \bf{1} & \bf{2}   	& \bf{3} 		& \bf{4} & \bf{5} & \bf{6} & \bf{7} & \bf{8} 	& \bf{9} 		\\   \hline
 $T$ 		& 2 		& 2 		& 2 		& 2 		& 2 		& 1.9 	& 1.6 	& 1.2 	& 0.87 	& 0.65 	\\
 $D$ 		& 0 		& 1.6 	& 1.5 	& 1.15 	& 0.9 	& 0.75 	& 0.65 	& 0.55 	& 0.52 	& 0.5		\\
 $K$ 		& 0 		& 0.2 	& 0.4 	& 0.6 	& 0.8 	& 0.9 	& 0.94 	& 0.96 	& 0.97 	& 0.98 	\\  \hline
\end{tabular} 
\captionof{table}{Parameter combinations defining the ten reference ARI's.}\label{table:tdk}
\end{center}

\vspace{0.5 cm}
Let $\hat V_j(t)$, where $j=0,\ldots,9$, denote the response of the model for the $j$th combinations of the parameters $(T,D,K)$. The ARI is determined by selecting the model producing the smallest root-mean-square (RMS) error:

\begin{equation}
\ari_{std}= \min_{j\in\{0,\ldots 9\}} \Big\|\frac{\hat V_j(t) - V[t]}{V_m}\Big\|. \label{ARI_standard}
\end{equation}

\noindent
The results of \ari are interpolated by cubic splines to include fractional values of the \ari. We use the subscript `standard' in the above definition of the ARI to distinguish it from the alternate definition we propose in the subsequent section.


\subsection{Bayesian inference}
We briefly describe the theory behind bayesian inference before applying it to our model. Assume we have a data set $X$ and a model for this data, $Y(\theta)$, which depends on some parameters  $\theta = \{\theta_1, \theta_2, .... \}$. In a traditional parameter fitting we would define a cost function $C(\theta) = \sum_j ||Y_j(\theta ) - X_j||$, with some suitable norm $|| . ||$ (The $L_2$ norm being the most popular). Then the best fit parameters $\theta^*$ are those which minimise this cost function: $\theta^* = \min_{\theta} C(\theta)$. 

How certain are we of $\theta^*$? Bayesian inference answers this question by providing a probability distribution $P(\theta | X)$ for the parameters $\theta$ given the data $X$, which quantifies this certainty. It can be computed using Bayes' theorem,

\begin{equation}
P(\theta | X) = \frac{P(X | \theta ) P(\theta)}{ P(X)}. \label{bayes}
\end{equation} 

\noindent
The term $P(\theta)$ is the prior distribution of the parameters $\theta$. This quantifies our belief of what the parameters $\theta$ could be, before observing the data $X$. If we have no prior beliefs, a uniform distribution is often used so that each parameter value is equally likely. The term $P(X | \theta) $ is the `likelihood', the probability of measuring the data $X$ given the parameter $\theta$. It is the asserted relationship between the model and the data. This is chosen by the modeller, and is dictated by the data in question. For parameter inference problems the likelihood is typically taken to be Gaussian. 

The last term $P(X)$ is often called the `evidence', and can be thought of as a normalisation constant. It can be expressed in terms of the prior and likelihood: $\int P(X | \theta) P(\theta) d \theta$. If we could compute this integral then we would have our desired distribution $P(\theta | X)$ via \eqref{bayes}. However, only in the simplest cases is this doable analytically, and so numerical approximations must be used.

\medskip

\noindent
\textbf{Markov Chain Monte Carlo}. There is a way to sample from the desired distribution $P(\theta | X)$ without ever computing the evidence $P(X)$. This is known as the Markov Chain Monte Carlo (MCMC) method, which we use in this work. There are many resources \cite{chib1995understanding,hastings1970monte,robert2004monte} which derive the theory behind MCMC, but for convenience, we give a brief description here, before returning to the ARI problem.

MCMC's basic idea is to construct a Markov chain whose limiting distribution is $P(\theta | X)$, our sought after distribution. A Markov chain is uniquely defined by a collection of states, $y$, and transition probabilities between these states, $P(y'|y)$, which do not depend on time (Note, in the context of bayesian parameter inference, these states correspond to points in the conditional parameter space $\theta | X$). At each time-step we move from a state $y$ to a new state $y'$ with probability $P(y'|y)$. As time evolves the Markov-chain is described by $P(y,t | y_0 )$, the probability to be in state $y$ at time $t$ having started in the state $y_0$. 

For certain transition probabilities $P(y'|y)$, there exists a stationary distribution 

\begin{equation}
\pi(y) = \lim_{t \rightarrow \infty} P(y,t | y_0). \label{stationary_distribution}
\end{equation}

\noindent
A sufficient condition for the existence of this state is 

\begin{equation}
P(y'|y) \pi(y) = P(y|y') \pi(y'). \label{detailed_balance}
\end{equation}

\noindent
The above condition is known as ``detailed balance", and can be thought of as a no net flux condition. Thus, if detailed balance can be satisfied \eqref{detailed_balance}, then Markov chain will have a stationary distribution $P(\theta | X)$, our goal. Our task is then to choose transition probabilities $P(y' |y)$ that satisfy \eqref{detailed_balance}. But how do we choose these?

\medskip

\noindent
\textbf{Metropolis-Hastings algorithm}.
Many different choices can be made, each of which corresponds to a different MCMC method. Common choices are the Metropolis-Hasting and Gibb's sampling algorithms. In this paper we use the Metropolis-Hastings algorithm, in which the transition probabilities are
\begin{equation}
P(y'|y) = g(y'|y) A(y'|y). \label{prob_decomp}
\end{equation}
The density $g(y'|y)$ is the \textit{proposal distribution}, and is typically taken to be a Gaussian. It proposes a transition to a new state $y'$. This proposal is then accepted or rejected with probability $A(y'|y)$, a quantity accordingly called the \textit{acceptance distribution}. Requiring detailed balance constrains $A(y' | y)$ as follows
\begin{equation}
\frac{A(y'|y)}{A(y'|y)} = \frac{P(y')}{P(y)} \frac{g(y'|y)}{g(y|y')}, \label{A_constraint}
\end{equation}
which we obtained by plugging \eqref{prob_decomp} into equation \eqref{stationary_distribution} for the stationary distribution $\pi(y)$. Thus our problem is reduced to choosing $A(y' | y)$ such that \eqref{A_constraint} is obeyed. The algorithm is then defined by the choice
\begin{equation}
A(y'|y) = \min \Big(1, \frac{\pi(y')}{\pi(y)} \frac{g(y'|y)}{g(y|y')} \Big) \label{metropolis_hastings}.
\end{equation}

We now translate the above results into the notation of the inference problem: the states of the Markov chain correspond to parameter values conditioned on our data $y \rightarrow \theta | X$, while the stationary distribution is the posterior distribution $\pi(y) \rightarrow P(\theta | X)$. Then,
\begin{equation}
A(y'|y) = \min \Big(1, \frac{P(X| \theta')}{P(X | \theta)} \frac{g(y'|y)}{g(y|y')} \Big) \label{metropolis_hastings}.
\end{equation}
Using Bayes' theorem \eqref{bayes} we can  replace the ratio $\frac{P(X| \theta')}{P(X | \theta)}$ with the right-hand side of \eqref{bayes} to obtain the final result

\begin{equation}
A(y'|y) = \min \Big(1, \frac{P(\theta' | X) P(\theta')}{P(\theta | X) P(\theta)} \frac{g(y'|y)}{g(y|y')} \Big) \label{metropolis_hastings}.
\end{equation}

\medskip

Although the theory behind MCMC is somewhat involved, using it as a tool is straightforward. The user simply specifies the likelihood $P(X|\theta)$ and a prior $P(\theta)$ as input, and then uses a software package to run the MCMC algorithm, which returns the desired posterior $P(\theta|X)$. There are many packages available for this purpose. We used python's ``PyMC". We recommend the e-book ``Bayesian methods for hackers" \cite{davidson2015bayesian} for a hands-on introduction to PyMC.



\section{Results}

\subsection{Alternate definition of ARI}
It is difficult to apply bayesian methods directly to the ARI as formulated above in \eqref{ARI_standard}. We thus propose an alternate definition of the \ari which does not have this drawback, and gives similar results as the original definition \eqref{ARI_standard}. Since this new \ari is calculated using an inverse method, we denote it as $\ari_{inv}$.

 The first step in calculating $\ari_{inv}$ is to fit $\hat{V}[t]$ to the known time series $V[t]$. Recall that $\hat{V}[t]$ is determined by the pressure time series $P[t]$, and the parameters $T,D,K$ 
\begin{equation}
\hat{V}_t = F (P[t], T,D,K).
\end{equation}
where $F(\cdot)$ represents equation \eqref{eq3} with equations \eqref{eq1} and \eqref{eq2} inserted. We find the best fit parameters $(T^*,D^*,K^*)$ by minimizing the following cost function
\begin{equation}
C(T,D,K) = \Big\| V_t - \hat{V}_t \Big\| ^2,
\label{cost_function}
\end{equation}
where $|| . ||^2$ denotes the $L^2$ norm. We then consider the quantity $V^*[t]$, which is the response of $\hat{V}[t]$ to a step function pressure drop, $H(t)$, at the best fit parameters,
\begin{equation}
V^*[t] := F \Big[ P(t) = H(t), T^*, D^*, K^* \Big ].
\label{V_tilde}
\end{equation}
$\ari_{inv}$ is then found by comparing the asymptotic value of $\tilde{V}$ to that of the reference flow rates $\hat{V}_j$
\begin{equation}
\ari_{inv} = \min_{j\in\{0,\ldots 9\}} \Big\|\frac{\hat V_j[t\rightarrow \infty] -  V^*[t\rightarrow \infty]}{V_m}\Big\|.
\label{ARI_inverse}
\end{equation}
As before the above expression can be extended to fractional values by appropriate scaling.

\begin{figure}[h!]
\centering
    \includegraphics[width= 1 \linewidth]{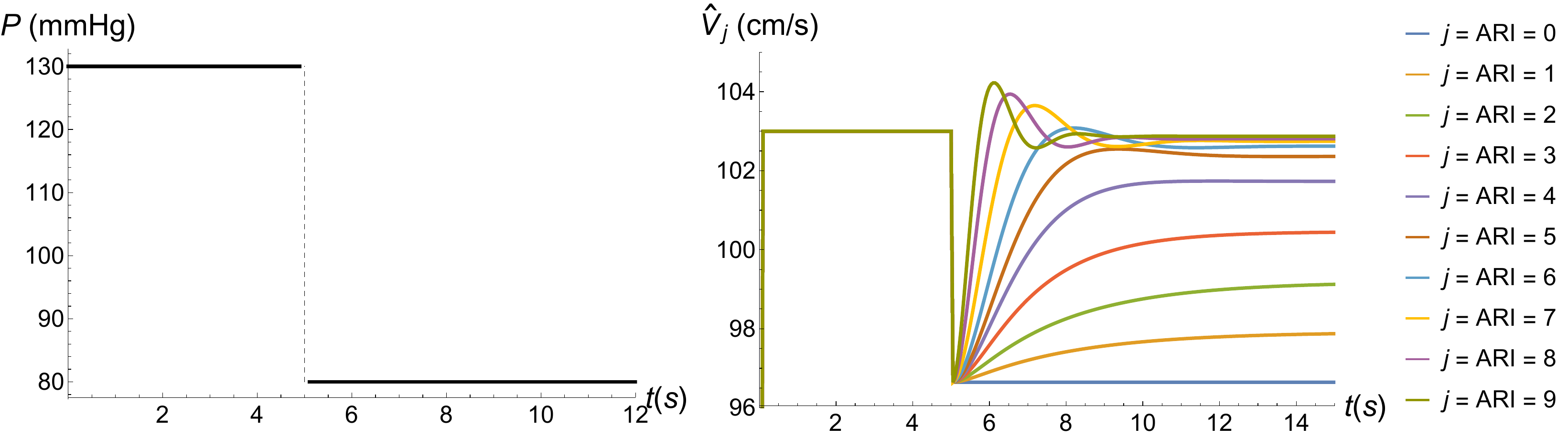}
  \caption{Response of reference cerebral blood flow rates, $\hat{V}_j$, to a step function drop in pressure. As shown in panel (a), the drop in pressure has magnitude $50$ mmHg and occurs at $t = 5$ s. Panel (b) shows the response of the reference blood flow rates $\hat{V}_j$ for $j = 0, \dots 9$ corresponding to an ARI of $0, \dots 9$ (see Table 1 for values for values of parameters $T,D,K$ which correspond to each ARI). As can be seen, $\hat{V}_9$ is temporally diminished after the drop in pressure at $t = 5$s, but eventually returns to its `pre-drop' value, indicating perfect auto-regulatory performance: $\ari=9$. On the other hand $\hat{V}_{0}$ shows no recovery after the drop in pressure, which corresponds to a lack of autoregulation: $\ari = 0$. The values between $0$ and $9$ then interpolate between these two extremes. }
  \label{schematics}
\end{figure}

The rationale behind defining $\ari_{inv}$ this way is illustrated in Figure~\ref{schematics}, where we show the response of the reference flow rates $\hat{V_j}$ to a step function drop in pressure. Recall that perfect autoregulation means the blood flow rate $\hat{V}$ remains constant in spite of changes in blood pressure. In the context of a step function drop in pressure, then, $\hat{V}$ should briefly change as the pressure drops, before returning to its original value. As can be seen, the reference flow rate $\hat{V_9}$ has this behaviour, correctly identifying that an ARI of $9$ denotes perfect autoregulation. Similarly, the remaining reference velocities $\hat{V}_{j < 9}$ which by definition have lower ARI's, asymptote to gradually diminished values. Thus, there is a correspondence between the asymptotic value of $\tilde{V}$ (in response to a step function drop in pressure) and the $\ari$, which motivates the definition of $\ari_{inv}$ given by \eqref{ARI_inverse}.



\subsection{Application of bayesian inference to ARI}
We now show how bayesian inference can be used to estimate the error in $\ari_{inv}$.  As described above, the definition of $\ari_{inv}$  \eqref{ARI_inverse} requires finding the best fit parameters $(T^*, D^*, K^*)$. Hence, a single parameter set $(T,D,K)$ corresponds to a single $\ari_{inv}$. A \textit{distribution} of parameter sets would thus correspond to a distribution of $\ari_{inv}$. We use bayesian inference for this purpose and find a distribution over the best fit parameters, $P(T,D,K | V)$ (conditioned on the data $V$), which quantifies our uncertainty in their values. In turn, the distribution $P(T,D,K | V)$ can be used to find a distribution of $\ari_{inv}$ values, whose width quantifies the error in $\ari_{inv}$.

We aim to find a distribution of parameters of best fit given our data set $P(T,D,K | V)$, which using Bayes' theorem, becomes
\begin{equation}
P(T,D,K| V) = \frac{P(V | T,D,K) \cdot P(T,D,K)}{ P(V)}.
\end{equation}
We made the following choices for the likelihood and prior
\begin{align}
P(V | T,D,K)        &= \exp \Big[ - {(V - \tilde{V} )}/{2 \sigma^2} \Big] \\
P(T,D,K, \sigma) &= \text{Uni}(0,2)\, \text{Uni}(0,1.0)\, \text{Uni} (0,1)\, \text{Gamma}(1,1).
\end{align}
We chose a Gaussian likelihood, which is equivalent to assuming the noise in our system is normally distributed with zero mean and standard deviation $\sigma$. There are two approaches to estimating $\sigma$. The first is that the modeler specifies an estimate from their knowledge of the physics of the problem. We were unsure how to do this for the blood flow rate time series $V$, since it has been pre-processed, which smoothes the noise. 

We therefore opted for the second approach, which is to treat $\sigma$ as an additional, unknown parameter to be estimated. For its prior, we chose the gamma distribution.  This is a two parameter distribution whose probability density function is $f(x)= \Gamma( \alpha) ^{-1} \beta^{\alpha} x^{\alpha - 1}  e^{- \beta x}$ for $x, \alpha, \beta  > 0 $ where $\Gamma$ is the gamma function. This is often used as a prior for a parameter whose scale is unknown, as is the case for  $\sigma$. The choices of $(\alpha, \beta) = (1,1)$ were arbitrary. To check they didn't influence the results, we ran our simulations for $\alpha = 0.1, 0.2, \dots 1.0$ and $\beta = 0.1, -0.2, \dots 1.0$, and found no significant differences in the computed values.

We took the priors on the remaining parameters to be uniform distributions on the interval $(a,b)$. The end points of the interval were chosen as the maximum and minimum values of the parameters in the reference ARI table, with the exception of $P(D)$, which has min and max values $0, 1.5$. We found that values of $1.0 < D < 1.5$ led to an oscillatory $\hat{V}$. To avoid this unphysical behaviour, we truncated the interval to $(0,1.0)$.

\begin{figure}[h!]
\centering
    \includegraphics[width=  \linewidth]{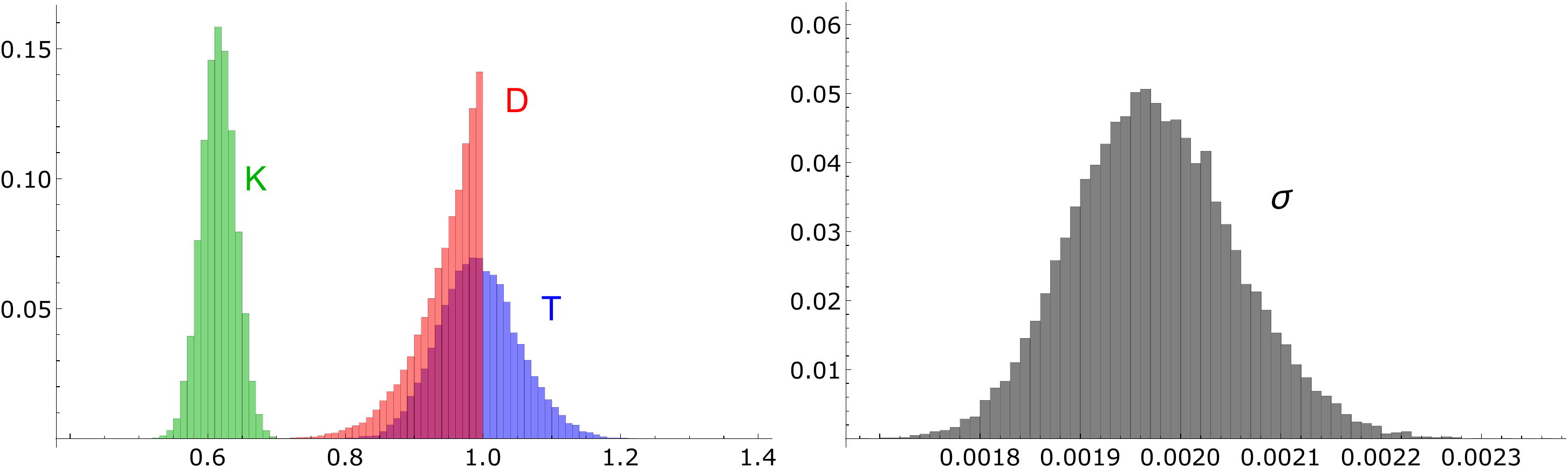}
  \caption{Posterior distributions of various parameters as computed by MCMC. $5 \times 10^5$ sample were taken, with a $50 \%$ burn-in and a thinning parameter of $10$. Left: Marginal distribution of parameters $T,D,K$. Right: Marginal distribution of noise parameter $\sigma$.}
  \label{posterior}
\end{figure}

Having specified the likelihood and priors we next computed the following posterior $P( T,D,K, \sigma |V )$. We took  $5 \times 10^5$ samples,  $50 \%$ burn-in, and a thinning parameter of 10 (these are numerical parameters required by PyMC). In Figure~\ref{posterior} we show the marginal distributions of each parameter. 

\begin{figure}[h!]
\centering
    \includegraphics[width= 0.65 \linewidth]{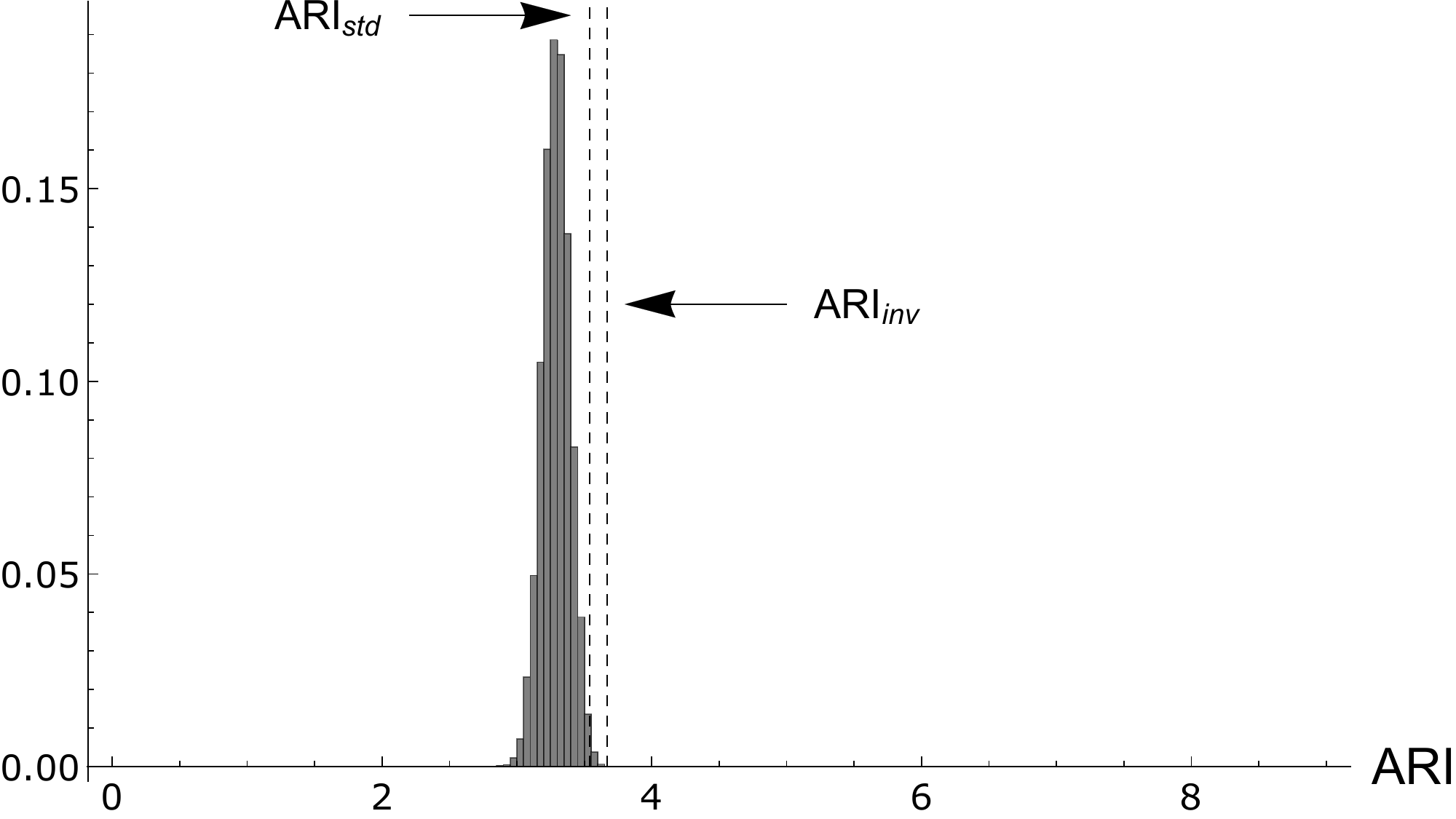}
  \caption{Distribution of ARIs resulting from the posterior distributions $P( T,D,K, \sigma |V )$. The ARI values computed using the inverse (equation \eqref{ARI_inverse}) and standard (equation \eqref{ARI_standard}) methods are also shown for reference.  }
  
  \label{posterior_ARI}
\end{figure}

We next computed a distribution of ARIs by plugging each triplet $(T,D,K)$ in the posterior distribution $P( T,D,K, \sigma |V )$ into equation \eqref{ARI_inverse} for $\ari_{inv}$. The result is shown in Figure~\ref{posterior_ARI}. A peaked distribution is evident, whose mean and width (defined as half the difference of the maximum and minimum values) finally give us the desired value and error of the ARI. We show the ARI computed from the standard and inverse methods for reference. As can be seen, the bayesian result in consistent with the others.

\begin{equation}
\begin{aligned}
\ari_{std} &= 3.7 \\
\ari_{inv} &= 3.5 \\
\ari_{bay} &= 3.4 \pm 0.3.
\end{aligned}
\end{equation}

\section{Discussion}
The ARI is used to assess dynamic CA performance. Previous methods computed a single ARI for a given patient. In this work, we showed how bayesian methods could be used to attach an error to a given ARI. We did this by using the Markov chain Monte Carlo method (in concert with an alternate definition of the ARI) to produce a probability distribution for a single ARI. We then used the width of this distribution to estimate the desired uncertainty in an ARI measurement. 

The connection between impaired CA and certain disorders have been widely dicussed in the literature. Certain brain disorders, such as stroke, subarachnoid haemorrhage and head injury appear to impair CA \cite{dawson2000dynamic,giller1990frequency,czosnyka1996monitoring}, which motivates the use of the ARI. Knowledge of the accuracy of an ARI reading, which our bayesian approach supplies, strengthens this potential. It gives a clinician a sense of the reliability in a given measurement.

The current work has some limitations, most notably in the way the data were collected. For example, the cerebral blood flow velocity was approximated by measuring the flow rate in the middle cerebral artery (MCA). This approximation is only valid if the diameter of the MCA is constant, which was unverified. Another limitation was in the length of the times series $V[t]$ and  $P[t]$, which was approximately two minutes. It is possible that this relatively short length is unrepresentative of the true behaviour of the ABP and CBFV, which would bias our results. 

Another problem with the time series is that they often contain various artifacts, both physiological and non-physiological. The sensitivity of the \ari to these artifacts is relatively unknown. Recent works, however, have started to explore this problem \cite{li2009artificial,mahdi2017effects}. The authors in \cite{mahdi2017effects} examined the response of the \ari to four common types of non-physiological artifact (as identified by Li et al in \cite{li2009artificial}), and found different qualitative effects. It would be interesting to see if our bayesian approach could corroborate their findings. That is, if the error estimates in the \ari in the presence of a given artifact would reflect the effects they reported.

One purpose of our work was to demonstrate the utility of bayesian statistics in biophysics problems. We chose perhaps the simplest example of how this could be done, namely by using Bayesian parameter
inference to estimate the error in ARI's. There many other more exotic bayesian tools which could used, which we hope will be explored in future work. On such contender is bayesian sequential analysis, a method which dynamically chooses when enough data have been collected in a sampling problem. This could be useful in determining, for example, how long time series of a patient's ABP and CBFV should be. At present, it is not clear what the optimal length is, although it is known that ABP time series with greater variance leads to more robust autoregulation indices, as described in \cite{mahdi2017increased}.

\section*{Acknowledgement}
A.M. acknowledges the partial support of the EPSRC project EP/K036157/1. K.P.O. acknowledges support from NSF grants DMS-1513179 and CCF-1522054. They also thank the MRC (expand). The authors thank Greg Mader, Katrina Johnson, and Erica Rutter for helpful discussions.


\begin{thebibliography}{10}

\bibitem{Aaslid1989}
R~Aaslid, K-F Lindegaard, W~Sorteberg, and H~Nornes.
\newblock Cerebral autoregulation dynamics in humans.
\newblock {\em Stroke}, 20(1):45--52, 1989.

\bibitem{chib1995understanding}
Siddhartha Chib and Edward Greenberg.
\newblock Understanding the metropolis-hastings algorithm.
\newblock {\em The american statistician}, 49(4):327--335, 1995.

\bibitem{Czosnyka2001}
M~Czosnyka, P~Smielewski, S~Piechnik, L~A Steiner, and JD~Pickard.
\newblock Cerebral autoregulation following head injury.
\newblock {\em Journal of neurosurgery}, 95(5):756--763, 2001.

\bibitem{czosnyka1996monitoring}
Marek Czosnyka, Piotr Smielewski, Peter Kirkpatrick, David~K Menon, and John~D
  Pickard.
\newblock Monitoring of cerebral autoregulation in head-injured patients.
\newblock {\em Stroke}, 27(10):1829--1834, 1996.

\bibitem{Dagal2009}
A~Dagal and AM~Lam.
\newblock Cerebral autoregulation and anesthesia.
\newblock {\em Current Opinion in Anesthesiology}, 22(5):547--552, 2009.

\bibitem{davidson2015bayesian}
Cameron Davidson-Pilon.
\newblock {\em Bayesian methods for hackers: probabilistic programming and
  Bayesian inference}.
\newblock Addison-Wesley Professional, 2015.

\bibitem{dawson2000dynamic}
Suzanne~L Dawson, Melanie~J Blake, Ronney~B Panerai, and John~F Potter.
\newblock Dynamic but not static cerebral autoregulation is impaired in acute
  ischaemic stroke.
\newblock {\em Cerebrovascular Diseases}, 10(2):126--132, 2000.

\bibitem{Dirnagl90}
U~Dirnagl and W~Pulsinelli.
\newblock Autoregulation of cerebral blood flow in experimental focal brain
  ischemia.
\newblock {\em J Cereb Blood Flow Metab.}, 10:327---336, 1990.

\bibitem{giller1990frequency}
Cole~A Giller.
\newblock The frequency-dependent behavior of cerebral autoregulation.
\newblock {\em Neurosurgery}, 27(3):362--368, 1990.

\bibitem{Harper84}
SL~Harper, HG~Bohlen, and MJ~Ruben.
\newblock Arterial and microvascular contributions to cerebral cortical
  autoregulation in rats.
\newblock {\em Am J Physiol Heart Circ Physiol.}, 246:H17---H24, 1979.

\bibitem{hastings1970monte}
W~Keith Hastings.
\newblock Monte carlo sampling methods using markov chains and their
  applications.
\newblock {\em Biometrika}, 57(1):97--109, 1970.

\bibitem{Lassen59}
NA~Lassen.
\newblock Cerebral blood flow and oxygen consumption in man.
\newblock {\em Physiol Rev.}, 39:183---238, 1959.

\bibitem{li2009artificial}
Qiao Li, Roger~G Mark, and Gari~D Clifford.
\newblock Artificial arterial blood pressure artifact models and an evaluation
  of a robust blood pressure and heart rate estimator.
\newblock {\em Biomedical engineering online}, 8(1):13, 2009.

\bibitem{Lip2000}
LA~Lipsitz, S~Mukai, J~Hamner, M~Gagnon, and V~Babikian.
\newblock Dynamic regulation of middle cerebral artery blood flow velocity in
  aging and hypertension.
\newblock {\em Stroke}, 31:1897--1903, 2000.

\bibitem{MacKenzie79}
ET~MacKenzie, JK~Farrar, W~Fitch, DI~Graham, PC~Gregory, and AM~Harper.
\newblock Effects of hemorrhagic hypotension on the cerebral circulation. i.
  cerebral bood flow and pial arteriolar caliber.
\newblock {\em Stroke.}, 10:711---718, 1979.

\bibitem{MacKenzie76}
ET~MacKenzie, S~Strandgaard, DI~Graham, JV~Jones, AM~Harper, and JK~Farrar.
\newblock Effects of acutely induced hypertension in cats on pial arteriolar
  caliber, local cerebral blood flow, and the blood-brain barrier.
\newblock {\em Circ Res.}, 39:33---41, 1976.

\bibitem{mader2015modeling}
Greg Mader, Mette Olufsen, and Adam Mahdi.
\newblock Modeling cerebral blood flow velocity during orthostatic stress.
\newblock {\em Annals of biomedical engineering}, 43(8):1748--1758, 2015.

\bibitem{mahdi2017increased}
Adam Mahdi, Dragana Nikolic, Anthony~A Birch, Mette~S Olufsen, Ronney~B
  Panerai, David~M Simpson, and Stephen~J Payne.
\newblock Increased blood pressure variability upon standing up improves
  reproducibility of cerebral autoregulation indices.
\newblock {\em arXiv preprint arXiv:1705.04942}, 2017.

\bibitem{mahdi2017effects}
Adam Mahdi, Erica~M Rutter, and Stephen~J Payne.
\newblock Effects of non-physiological blood pressure artefacts on cerebral
  autoregulation.
\newblock {\em Medical Engineering \& Physics}, 2017.

\bibitem{Panerai08b}
RB~Panerai.
\newblock Cerebral autoregulation: from models to clinical applications.
\newblock {\em Cardiovasc Eng.}, 8:43--59, 2008.

\bibitem{Panerai08}
RB~Panerai, EL~Sammons, SM~Smith, WE~Rathbone, S~Bentley, JF~Potter, and
  NJ~Samani.
\newblock Continuous estimates of dynamic cerebral autoregulation: influence of
  non-invasive arterial blood pressure measurements.
\newblock {\em Physiol Meas.}, 29:497--513, 2008.

\bibitem{Panerai2000}
RB~Panerai, DM~Simpson, ST~Deverson, P~Mahony, P~Hayes, and DH~Evans.
\newblock Multivariate dynamic analysis of cerebral blood flow regulation in
  humans.
\newblock {\em Biomedical Engineering, IEEE Transactions on}, 47(3):419--423,
  2000.

\bibitem{robert2004monte}
Christian~P Robert.
\newblock {\em Monte carlo methods}.
\newblock Wiley Online Library, 2004.

\bibitem{Tiecks95}
FP~Tiecks, AM~Lam, R~Aaslid, and DW~Newell.
\newblock Comparison of static and dynamic cerebral autoregulation
  measurements.
\newblock {\em Stroke.}, 26:1014---1019, 1995.

\bibitem{Zong2003}
W~Zong, T~Heldt, GB~Moody, and RG~Mark.
\newblock An open-source algorithm to detect onset of arterial blood pressure
  pulses.
\newblock In {\em Computers in Cardiology, 2003}, pages 259--262. IEEE, 2003.

\end{thebibliography}
\end{document}